\documentclass[3p,times,twocolumn]{elsarticle}
\usepackage{ecrc}

\volume{00}
\firstpage{1}
\journalname{Nuclear Physics B Proceedings Supplement}

\runauth{B.-H. Lee and D. Yeom}
\jid{nuphbp}
\jnltitlelogo{Nuclear Physics B Proceedings Supplement}

\usepackage{amssymb}
\usepackage{amsthm}
\usepackage[figuresright]{rotating}

\begin{document}

\begin{frontmatter}

\dochead{}

\title{Status report: black hole complementarity controversy\fntext[fn1]{A proceeding for the 9th International Symposium on Cosmology and Particle Astrophysics (CosPA2012). Talk on the 14th of November, 2012, Taipei, Taiwan.}}


\author{Bum-Hoon Lee}
\ead{bhl@sogang.ac.kr}

\author{Dong-han Yeom}
\ead{innocent.yeom@gmail.com}

\address{Center for Quantum Spacetime, Sogang University, Seoul 121-742, Republic of Korea}

\begin{abstract}
Black hole complementarity was a consensus among string theorists for the interpretation of the information loss problem. However, recently some authors find inconsistency of black hole complementarity: large $N$ rescaling and AMPS argument. According to AMPS, the horizon should be a firewall so that one cannot penetrate there for consistency. There are some controversial discussions on the firewall. Apart from these papers, the authors suggest an assertion using a semi-regular black hole model and we conclude that the firewall, if it exists, should affect to asymptotic observer. In addition, if any opinion does not consider the duplication experiment and the large $N$ rescaling, then the argument is difficult to accept.
\end{abstract}

\begin{keyword}
black hole information loss problem, black hole complementarity, regular black hole
\end{keyword}

\end{frontmatter}

\section{Introduction}

Black hole complementarity \cite{Susskind:1993if} is an interpretation to understand the information loss problem in black hole physics \cite{Hawking:1976ra}. This is motivated from our beliefs on the natural laws. First, we hope to believe the \textit{unitarity} of quantum mechanics; for any observer, the sum of all possible probabilities should be unity and it should not be smaller or larger than one. This implies that information should be conserved and the nature does not allow to observe the loss or copy of information. Second, we hope to believe the \textit{semi-classical description} of a black hole by using the local quantum field theory for an observer outside the event horizon. The semi-classical calculations (Hawking temperature, evaporation of black holes, etc.) should be a good description unless we consider the singularity. Third, we hope to believe that \textit{general relativity} should be a good description for an in-going observer inside the event horizon.

\begin{figure}
\begin{center}
\includegraphics[scale=0.7]{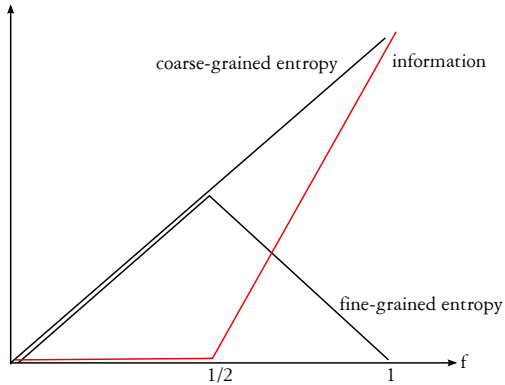}
\caption{\label{fig:information_retention}Information emission from a black hole.}
\end{center}
\end{figure}

Now the question is whether these three assumptions are consistent or not. In this context, Page \cite{Page:1993df} shows an interesting discussion on the black hole information issue. For a given closed system $U$, we can divide this to two subsystems: $A$ and $B$. Here, $A$ is interpreted as a black hole and $B$ is interpreted as the background. The number of states of $A$ is $n$ and the number of states of $B$ is $m$. The entire system is closed so that $n \times m$ is a constant, while $n$ and $m$ can vary. Initially, $m = 1$ and, as the black hole evaporates, $n$ decreases and eventually approaches to $1$ when the evaporation ends. Now, the \textit{mutual information} between $A$ and $B$ is defined as follows:
\begin{eqnarray}
I(B:A) = S(B) - S(B\;|\;A),
\end{eqnarray}
where $S(B) = \log m$ is called by the \textit{coarse-grained entropy} and $S(B\;|\;A)$ is called by the \textit{fine-grained entropy}, or the entanglement entropy. From the estimation of Page, for a given pure and random state, all the mutual information should be transferred from $A$ to $B$ and $A$ begins to send a bit of information to $B$ when the black hole entropy (coarse-grained entropy) decreased its half value (Figure~\ref{fig:information_retention}).

If we further assume that the black hole area is proportional to the logarithm of the number of states, then the information should be emitted after the black hole area decreased to the half value. This time scale is on the order of the lifetime $\sim M^{3}$ of the black hole and this time is called the \textit{information retention time}. One note is that the black hole can be still semi-classical even after the information retention time. Therefore, the only way to take out a bit of information is to rely on Hawking radiation: hence, \textit{Hawking radiation should have information}.

Then, are these assumptions/results self-consistent? In literature, we can list three important historical stages on this issue. First, Susskind and Thorlacius \cite{Susskind:1993mu} considered the consistency of black hole complementarity by considering the duplication experiment. In addition, even though we generalize black hole complementarity to the \textit{scrambling time} \cite{Hayden:2007cs}, this principle seemed to be viable. Second, some authors discussed that semi-classical black holes allow the duplication experiment when we assume a large number of scalar fields. Dvali \cite{Dvali:2007hz} considered this problem, but the considered examples were not semi-classical. Yeom and Zoe \cite{Yeom:2009zp} considered rather semi-classical black holes with large $N$ rescaling and could confirm that a large number of scalar fields can allow the duplication experiment \cite{Yeom:2008qw}. Third, recently, Almheiri, Marolf, Polchinski and Sully \cite{Almheiri:2012rt} discussed that black hole complementarity in itself is not consistent from a field theoretical argument. After the paper of AMPS, this issue is beginning to be focused by literature \cite{Susskind:2012rm,Mathur:2012jk,Bousso:2012as,Ori:2012jx,Hwang:2012nn}.

In this paper, we summarize these controversy and show perspectives for future studies. In Section~\ref{sec:dup}, we summarize the duplication experiment and the consistency check for black hole complementarity. In Section~\ref{sec:lar}, we show that the large $N$ rescaling can be used to allow the duplication experiment for any dimensions $D\geq3$. In Section~\ref{sec:fir}, we comment on the recent suggestion by AMPS on the firewall and the firewall controversy. Finally, in Section~\ref{sec:per}, we summarize and illustrate future perspectives.

\section{\label{sec:dup}Duplication experiment}

If the assumptions of black hole complementarity are true at the same time, then it seems contradictory because there are two copies of information: one is inside the horizon and the other is outside the horizon by Hawking radiation. However, if there is no observer who can see the both of the copied information at the same time, then black hole complementarity can be still safe; this can be similar with the case of particle-wave complementarity.

\begin{figure}
\begin{center}
\includegraphics[scale=0.7]{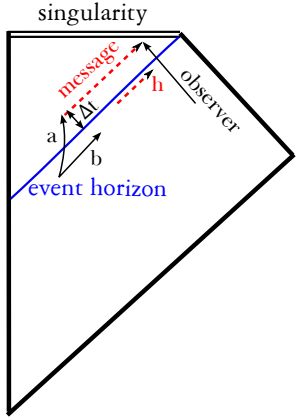}
\caption{\label{fig:Schwarzschild_duplication}Conceptual description of duplication experiment. Alice sends a message and Bob is the observer who measures $h$ and the message of Alice.}
\end{center}
\end{figure}

Let us define the \textit{duplication experiment} and check whether it is allowed or not in principle (Figure~\ref{fig:Schwarzschild_duplication}). We illustrate this experiment more technically. Step 1: Create an entangled spin pare $a$ and $b$. Step 2: An observer Alice falls into the black hole with the spin $a$. However, $b$ is still outside. Step 3: Alice sends a signal on $a$ along the out-going direction before she touches the singularity. Step 4: There is an observer Bob who is outside the event horizon. Bob first measures the state of $b$ and he knows what it is. Second, Bob waits and measures Hawking radiation outside the event horizon and fortunately measures the information of $a$ that is attached by Hawking radiation after the information retention time. In principle, Bob can know that a Hawking particle contains the information of $a$ by comparing with the state of $b$. We call this information $h$. Step 5: Bob falls into the black hole and fortunately sees the signal of Alice. Step 6: Then Bob knows that he has two copied quantum states and copied quantum information that is definitely violates the unitarity principle.

We can carefully illustrate the assumptions that we used for this experiment. First of all, we used three assumptions of black hole complementarity as we commented in Introduction. Second, we assumed the \textit{area-entropy relation}: then the information retention time is $\sim M^{3}$ and Hawking radiation should contain information. Finally, to justify Step 4 and Step 5, we implicitly assume that \textit{there is an observer who can read a bit of information from Hawking radiation}. If one of these assumptions is not satisfied, then the duplication experiment cannot success. On the other hand, if we assume these assumptions, then the duplication experiment is well-defined in principle.

Now let us check whether this is indeed possible or not. A black hole has the spatial size $r_{0} \sim M$ and Bob falls into the black hole after the time $\tau \sim M^{3}$. It is estimated that Alice should send a signal to Bob during the time $\Delta t \sim \exp -\tau/r_{0}$ for Step 3 \cite{Susskind:1993mu}. Because of the uncertainty relation, to send a bit of information, one should use the energy $\Delta E \sim 1/\Delta t$. If $\Delta E > M$, then such event seems to improbable to happen in realistic cases. Therefore, if $\tau > M \log M$, then it seems that the observation of the duplication of information is impossible. In fact, if a black hole works as a fast scrambler \cite{Hayden:2007cs}, then information can be escaped after the time $M \log M$. However, even in this marginal case, still the relation $\Delta E > M$ holds and hence black hole complementarity seems to be safe.

\section{\label{sec:lar}Large $N$ rescaling}

However, some authors suggested that the duplication experiment is possible if we consider a large number of scalar fields which contribute to Hawking radiation \cite{Yeom:2009zp,Yeom:2008qw}. This is related to the parameter rescaling. Let us first consider the semi-classical equation:
\begin{eqnarray}
G_{ab} = 8\pi \left( T_{ab} + N\hbar \langle T_{ab}\rangle \right).
\end{eqnarray}
By fixing $G = c = N\hbar = 1$, we can obtain a solution of the equation. However, it is not the results in the Planck units. When we change the results to the Planck units, we need the parameter rescaling. The transform rule is simple: if a quantity $X$ has a length dimension $[X]=L^{\alpha}$, then the following quantity is the value of $X$ in Planck units with large $N$:
\begin{eqnarray}
X' = X N^{\alpha/(D-2)},
\end{eqnarray}
where $D$ is the space-time dimension.

Then the duplication experiment should be reconsidered. The time difference and the mass should be rescaled in the large $N$ limit:
\begin{eqnarray}
\Delta t &\rightarrow& N^{1/(D-2)} \Delta t,\\
M &\rightarrow& N^{(D-3)/(D-2)} M.
\end{eqnarray}
Therefore, the required energy to send a signal to Bob is
\begin{eqnarray}
\Delta E = N^{-1/(D-2)} \frac{1}{\Delta t}.
\end{eqnarray}
Then, the consistency condition is
\begin{eqnarray}
N^{-1/(D-2)} \frac{1}{\Delta t} > N^{(D-3)/(D-2)} M.
\end{eqnarray}

Let us apply for the scrambling time. The scrambling time is $\tau \sim T^{-1} \log S$ where $T$ is the Hawking temperature and $S$ is the entropy. Therefore, it is on the order of $\sim r_{0} \log r_{0}$ in general. After the rescaling, we obtain $\tau/r_{0} \sim \log N^{1/(D-2)} r_{0}$. Therefore, if
\begin{eqnarray}
N > \frac{1}{M\Delta t} \sim \frac{r_{0}}{M} N^{1/(D-2)},
\end{eqnarray}
then the observation of the duplication is allowed. When we consider a sufficient number of scalar fields (does not necessarily excessively large), the observation of the duplication of information is possible \cite{Yeom:2009zp,Yeom:2008qw,Hwang:2012nn}.

\begin{figure}
\begin{center}
\includegraphics[scale=0.6]{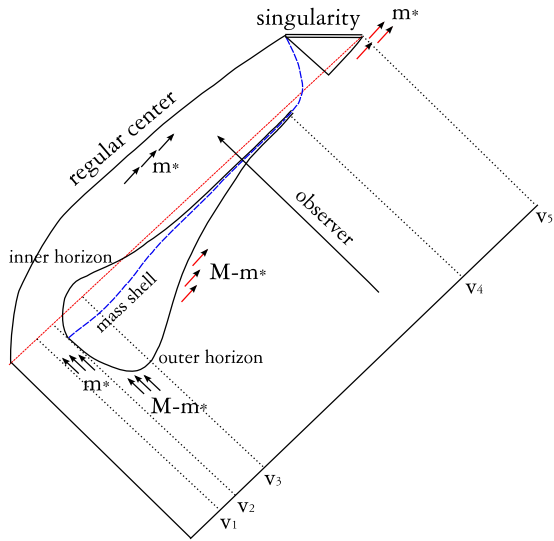}
\caption{\label{fig:regular}A causal structure of an evaporating regular black hole. This model is suggested in \cite{Yeom:2008qw} and confirmed by a numerical simulation in \cite{Hwang:2012nn}. Between the time $v_{1}$ and $v_{3}$ matter is collapsed. Due to the false vacuum near the center, the formation of the singularity is postponed and two horizons appear. After the evaporation, the black hole eventually form a singularity. However, the duplication experiment is well defined after the information retention time.}
\end{center}
\end{figure}

When we just consider the information retention time, the required number of scalar fields (in $4$-dimensions) to violate black hole complementarity is on the order of $\sim \exp M^{2}$. This is a quite large number. For a charged black hole or a regular black hole (Figure~\ref{fig:regular}), the duplication experiment can be done inside the inner apparent horizon \cite{Yeom:2008qw}. In general, the inner apparent horizon has mass inflation and to regulate such instability, we need a large number of scalar fields again, on the order of $\sim \exp \kappa_{i} v \sim \exp M^{2}$, where $\kappa_{i}$ is the surface gravity of the inner horizon and $v$ is the advanced time coordinate. Therefore, regarding the information retention time, the required number of scalar fields is quite large; although there is no fundamental limitation to assume the bound of scalar fields, it is also fair to say that one can still have a doubt for this large $N$ rescaling.

However, recently, one important progress was done: in two dimensions \cite{2D}, we should define a new rescaling law and this allows the duplication experiment (regarding the information retention time) without assuming the exponentially large number of scalar fields. Therefore, to summarize, we can conclude that there is \textit{no} doubt for the possibility of the duplication experiment with a reasonable (not excessively large) number of scalar fields.

\begin{figure}
\begin{center}
\includegraphics[scale=0.7]{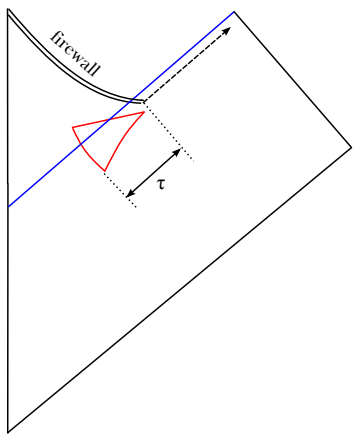}
\caption{\label{fig:wall}We can build a combination that, after the information retention time, two horizons disappear as in Figure~\ref{fig:regular}. However, the firewall should grow to prevent the duplication experiment. Now the firewall is outside the event horizon and there is no screen of the effects from the naked firewall.}
\end{center}
\end{figure}

\section{\label{sec:fir}Review of the firewall controversy}

The inconsistency of the original version of black hole complementarity is accepted by many authors \cite{Susskind:2012rm}. One of possible resolution is to drop the general relativity for an in-falling observer. This was introduced by AMPS; they said that an in-falling observer should see a dramatic event around the horizon due to a \textit{firewall} \cite{Almheiri:2012rt}. The firewall should be a kind of singularity around the horizon and it seems to grow from the central singularity to the horizon size singularity during the information retention time \cite{Susskind:2012rm}.

However, the exact location of the firewall is controversial. In fact, the Hawking radiation generating surface (apparent horizon) is outside the event horizon. Alice can send a signal to Bob between the apparent horizon and the event horizon; via the large $N$ rescaling, we can make this duplication possible in principle \cite{Hwang:2012nn}. Then, to prevent such a duplication experiment, the firewall should be outside the event horizon. This phenomena can be seen drastically for a regular black hole (Figure~\ref{fig:wall}). Then no one can prevent to see the effects of the firewall singularity.

There are interesting controversy in literature. We can ask three critical questions and group the opinions of the authors.
\begin{itemize}
\item[Q1.] Is the construction of the duplication experiment possible (probably using large $N$ rescaling)?
\item[Q2.] Is the inconsistency argument of AMPS correct?
\item[Q3.] If black hole complementarity is inconsistent, then do you believe there is a firewall?
\end{itemize}

If one says yes for Q1, Q2, and Q3 at the same time, then the firewall should be outside the event horizon and hence one should say that the asymptotic observer should see effects from the firewall \cite{Hwang:2012nn}. The firewall should completely prevent a free-infall.

If one says no for Q1 but yes for Q2 and Q3, then the fuzzball picture is also a possible option \cite{Mathur:2012jk}. The fuzzball picture allows free-infall when a probe has a sufficient energy. However, such papers should be sure whether one can surely say no for Q1 or not.

Also, many papers were involved to say no for Q2 and hence no for Q3 \cite{Bousso:2012as}. However, if one does not answer against the question Q1, then these papers are less attractive.

If one says no for Q3 whatever Q1 and Q2 \cite{Ori:2012jx}, then one has to drop other assumption for example unitarity or the area-entropy relation.

\section{\label{sec:per}Perspectives}

There is still an active discussion on the meaning and existence of the firewall. However, in many contexts, the large $N$ rescaling is ignored. At once if we consider the large $N$ rescaling in detail \cite{Yeom:2009zp}, we should say yes for Q1 and Q2. The fuzzball picture \cite{Mathur:2012jk} in itself is not easy to accept, since the free-infall allows the duplication experiment. Also, even though some arguments can resolve the problem of AMPS \cite{Bousso:2012as}, it is not sufficient to say that the original version of black hole complementarity is safe. Moreover, the large $N$ rescaling strongly suggests that the firewall is, if it exists, outside the event horizon \cite{Hwang:2012nn}. Then, we may see very large scale (horizon-sized) quantum gravitational effects, or the firewall is inconsistent. As a personal opinion, we carefully suggest that this may indirectly shows another possibility: the other assumption, e.g., the area-entropy relation, is inconsistent. For more firm conclusion, we have to develop our theoretical techniques further.

\section*{Acknowledgment}

The authors would like to thank the hospitality we received during the
9th International Symposium on Cosmology and Particle Astrophysics (CosPA2012).
The authors would like to thank Wontae Kim and Dong-il Hwang for helpful discussions. The authors are supported by the National Research Foundation of Korea(NRF) funded by the Korea government(MEST, 2005-0049409) through the Center for Quantum Spacetime(CQUeST) of Sogang University.

\nocite{*}
\bibliographystyle{elsarticle-num}
\bibliography{martin}

\end{document}